\documentclass{article} 
\usepackage{iclr2025_conference,times}

\usepackage{amsmath,amsfonts,bm}

\def\eqref#1{equation~\ref{#1}}

\def\1{\bm{1}}

\DeclareMathAlphabet{\mathsfit}{\encodingdefault}{\sfdefault}{m}{sl}
\SetMathAlphabet{\mathsfit}{bold}{\encodingdefault}{\sfdefault}{bx}{n}

\usepackage{hyperref}
\usepackage{url}
\usepackage{graphicx}

\title{A collection of innovations in Medical AI for patient records in 2024 }

\author{Yuanyun Zhang  \\
Department of Computer Science\\
University of the Chinese Academy of Sciences \\
\texttt{\{yuanyun81\}@ucas.ac.cn}
\And
Shi Li \\
Department of Computer Science \\
Columbia University \\
}

\iclrfinalcopy 
\begin{document}

\maketitle

\begin{abstract}

The field of Artificial Intelligence (AI) in healthcare is evolving at an unprecedented pace, driven by rapid advancements in machine learning and the recent breakthroughs in large language models (LLMs) \citep{zhao2023survey}. While these innovations hold immense potential to transform clinical decision-making, diagnostics, and patient care, the accelerating speed of AI development has outpaced traditional academic publishing cycles. As a result, many scholarly contributions quickly become outdated, failing to capture the latest state-of-the-art methodologies and their real-world implications. This paper advocates for a new category of academic publications—an annualized citation framework—that prioritizes the most recent AI-driven healthcare innovations. By systematically referencing the breakthroughs of the year, such papers would ensure that research remains current, fostering a more adaptive and informed discourse. This approach not only enhances the relevance of AI research in healthcare but also provides a more accurate reflection of the field’s ongoing evolution.
\end{abstract}

\section{Introduction}

Artificial Intelligence (AI) has become an integral force in shaping the future of healthcare \citep{rajpurkar2022ai}, with applications spanning from predictive modeling and diagnostics \citep{kasula2021ai} to personalized medicine and automated clinical decision support. Recent advancements in deep learning, particularly the rise of large language models (LLMs) \citep{zhao2023survey}, have further accelerated innovation, unlocking new possibilities for applying large scale models to biomedical research. However, this rapid progress presents a fundamental challenge: the traditional academic publishing cycle struggles to keep pace with the speed of AI development. By the time a study is peer-reviewed and published, newer, more advanced models and techniques may have already rendered its findings outdated or incomplete.

This discrepancy raises critical concerns about the relevance and longevity of AI research in healthcare. While conventional papers provide valuable theoretical and empirical contributions, they often fail to reflect the field's most current state. As AI-driven solutions continuously redefine the boundaries of what is possible, there is a growing need for an alternative publication model that prioritizes recent innovations.

To address this issue, we propose a new category of academic papers that explicitly cite the breakthroughs of the year, ensuring that discussions and analyses remain grounded in the latest developments. This approach would not only improve the timeliness of AI research in healthcare but also foster a more dynamic and iterative academic ecosystem. By incorporating up-to-date references and acknowledging the fluid nature of AI advancements, such papers would enhance the accuracy, relevance, and practical impact of AI-driven healthcare research.

\section{Language Models}


In 2024, the field of biomedical natural language processing (NLP) witnessed significant advancements with the development of several specialized language models tailored for health and biomedical applications. These models have been designed to enhance various tasks, including information retrieval, question answering, and text generation within the biomedical domain.

One notable contribution is MediSwift, introduced by \cite{thangarasa-etal-2024-mediswift}. This model employs sparse pre-training techniques on domain-specific biomedical text data, achieving up to 75\% weight sparsity during pre-training. This approach results in a 2-2.5x reduction in training FLOPs, leading to more computationally efficient models without compromising performance. The study demonstrates that sparse pre-training, combined with dense fine-tuning and soft prompting, offers an effective method for creating high-performing, computationally efficient models in specialized domains. 

In the realm of biomedical text retrieval, BMRetriever \citep{xu-etal-2024-bmretriever} was developed to enhance retrieval tasks across various biomedical applications. This model series utilizes unsupervised pre-training on large biomedical corpora, followed by instruction fine-tuning on a combination of labeled datasets and synthetic pairs. Experiments across multiple biomedical tasks and datasets have verified BMRetriever's efficacy, demonstrating strong parameter efficiency. Notably, the 410M variant outperforms baselines up to 11.7 times larger, and the 2B variant matches the performance of models with over 5B parameters. 

Addressing the challenge of referring ability in biomedical language models, \cite{jiang-etal-2024-improving} introduced a method to improve this aspect by designing a pre-training sequence that enhances the model's capacity to refer to entities within biomedical texts. Empirical studies demonstrate that this approach improves both intra-sample and inter-sample referring abilities of auto-regressive language models in the biomedical domain, encouraging more profound consideration of task-specific pre-training sequence design for continual pre-training. 

Another significant development is BioMedLM \citep{bolton2024biomedlm27bparameterlanguage}, a 2.7 billion parameter GPT-style autoregressive model trained exclusively on PubMed abstracts and full articles. Despite its relatively smaller size compared to models like GPT-4 \citep{achiam2023gpt} and Med-PaLM 2 \citep{singhal2025toward}, BioMedLM demonstrates competitive performance on multiple-choice biomedical question-answering tasks. For instance, it achieves a score of 57.3\% on MedMCQA (dev) \citep{pal2022medmcqa} and 69.0\% on the MMLU Medical Genetics exam \citep{wang2024mmlu}. This indicates that smaller, targeted models can serve as transparent, privacy-preserving, and economical foundations for specific NLP applications in biomedicine. 

Additionally, BioMistral \citep{labrak-etal-2024-biomistral} was introduced as an open-source large language model tailored for the biomedical domain. Utilizing Mistral as its foundation model and further pre-trained on PubMed Central, BioMistral has been evaluated on a benchmark comprising 10 established medical question-answering tasks in English, demonstrating its applicability in health contexts.

\section{EHR Foundation Models}

\begin{table*}[h]
    \centering
    \caption{Summary of Foundation Models for Electronic Health Records (EHRs) in 2024}
    \label{tab:foundation_models}
    \begin{tabular}{|p{3cm}|p{4cm}|p{6cm}|}
        \hline
        \textbf{Model} & \textbf{Key Features} & \textbf{Applications} \\
        \hline
        \textbf{EHRMamba} \citep{fallahpour2024ehrmambageneralizablescalablefoundation} & Scalable architecture using Mamba, supports long sequences, Multitask Prompted Finetuning (MTF), HL7 FHIR standard & General-purpose EHR model, clinical forecasting, multi-task learning in healthcare \\
        \hline
        \textbf{MOTOR} \citep{steinberg2023motor} & Time-to-event foundation model, explicit modeling of event likelihood and timing & Risk stratification, chronic disease management, ICU monitoring \\
        \hline
        \textbf{Context Clues} \citep{wornow2024contextcluesevaluatinglong} & Evaluates long-context models for clinical prediction tasks, balances computational cost with predictive accuracy & Long-term patient record analysis, temporal modeling in EHRs \\
        \hline
        \textbf{CORE BEHRT (Optimized)} \citep{odgaard2024corebehrtcarefullyoptimizedrigorously} & Fine-tuned and optimized version of BEHRT for EHR data, rigorous evaluation & Disease prediction, patient trajectory modeling \\
        \hline
        \textbf{CEHR-GPT} \citep{pang2024cehrgptgeneratingelectronichealth} & Clinical text generation using large-scale pre-training on medical text data & Automated report generation, clinical documentation, discharge summaries \\
        \hline
        \textbf{Event Stream GPT} \citep{mcdermott2023eventstreamgptdata} & GPT-based model capturing sequential dependencies in EHR data & Disease progression modeling, treatment pathway optimization \\
        \hline
        \textbf{Retrieval-Enhanced Medical Prediction} \citep{kim2024generalpurposeretrievalenhancedmedicalprediction} & Integrates retrieval mechanisms to dynamically retrieve past patient events & Context-aware clinical decision support, lifelong patient record modeling \\
        \hline
        \textbf{MEME} \citep{lee2024multimodal} & Converts multimodal EHR data into pseudo-notes, uses embedding models for modality separation & Emergency department decision support, multimodal learning for EHRs \\
        \hline
        \textbf{MetaGP} \citep{liumetagp} & 13-billion-parameter generative model integrating EHR data and medical literature & Rare disease diagnosis, emergency condition management, biomedical research \\
        \hline
        \textbf{EHRAgent} \citep{shi2024ehragent} & LLM-powered agent with code interface for autonomous data analysis & Few-shot learning for EHR reasoning, multi-tabular medical problem solving \\
        \hline
    \end{tabular}
\end{table*}

In 2024, foundation models for Electronic Health Records (EHRs) have seen remarkable advancements, addressing long-standing challenges such as computational efficiency, handling long patient histories, and adapting across diverse clinical tasks \citep{evans2016electronic}. As EHRs continue to serve as a vital source of patient information, these new models are pushing the boundaries of predictive accuracy and clinical decision-making by better capturing the complex, heterogeneous nature of medical data.

One of the key innovations in this space is the push for more scalable architectures that can handle longer patient histories without sacrificing computational efficiency. EHRMamba \citep{fallahpour2024ehrmambageneralizablescalablefoundation} exemplifies this trend by leveraging the Mamba architecture, which allows it to process sequences up to four times longer than traditional models while maintaining linear computational complexity. By introducing Multitask Prompted Finetuning (MTF), EHRMamba is capable of learning multiple clinical tasks simultaneously, making it an efficient choice for deployment across different healthcare applications. Its use of the HL7 FHIR data standard further enhances its interoperability with existing hospital systems, making it easier to integrate into real-world clinical workflows. Benchmarks on the MIMIC-IV dataset \citep{johnson2020mimic} confirm its strong performance across six key clinical prediction tasks, positioning it as a state-of-the-art model in EHR forecasting.

Beyond general-purpose EHR models, several efforts have focused on temporal modeling, as predicting the timing of future medical events is critical in clinical practice. MOTOR \citep{steinberg2023motor} is one such time-to-event foundation model designed to improve risk stratification and proactive patient management. By explicitly modeling the likelihood and timing of future clinical events, MOTOR enhances real-time monitoring capabilities, making it particularly valuable for chronic disease management and intensive care settings. This focus on long-term patient history is further explored in Context Clues: Evaluating Long Context Models for Clinical Prediction Tasks on EHRs \citep{wornow2024contextcluesevaluatinglong}, which examines how well different architectures handle extended patient records. The study highlights both the benefits and limitations of incorporating long-term data, providing insights into how models should be optimized to balance computational cost with predictive accuracy.

Another key area of improvement has been the refinement of existing architectures to better fit EHR-specific challenges. Building on the BEHRT framework \citep{li2020behrt}, A Carefully Optimized and Rigorously Evaluated BEHRT \citep{odgaard2024corebehrtcarefullyoptimizedrigorously} presents an optimized version of this transformer-based model. The study demonstrates how careful fine-tuning and rigorous evaluation lead to superior performance across multiple clinical tasks, emphasizing the importance of domain-specific adjustments when adapting foundation models for healthcare.

While many models focus on structured EHR data, there has also been an increasing interest in bridging the gap between structured and unstructured clinical notes. CEHR-GPT \citep{pang2024cehrgptgeneratingelectronichealth}, for instance, is designed to generate high-quality clinical text, automating documentation tasks such as discharge summaries and radiology reports. By training on a diverse corpus of clinical narratives, CEHR-GPT produces more coherent and contextually accurate reports, reducing the documentation burden on healthcare professionals and improving standardization across medical records.

In parallel, models like Event Stream GPT \citep{mcdermott2023eventstreamgptdata} have focused on capturing the sequential nature of clinical events. By leveraging a GPT-based architecture, this model learns dependencies between events in a patient’s medical history, aiding in disease progression modeling and personalized treatment planning. This approach allows for a deeper understanding of patient trajectories and could prove useful in complex conditions where interactions between multiple factors play a crucial role in disease evolution.

Another promising trend is the integration of retrieval mechanisms to enhance medical predictions. The General-Purpose Retrieval-Enhanced Medical Prediction Model Using Near-Infinite History \citep{kim2024generalpurposeretrievalenhancedmedicalprediction} exemplifies this direction by leveraging extensive patient histories to inform clinical predictions. Unlike conventional models that process patient data sequentially, this model retrieves relevant past events dynamically, ensuring that long-term patterns are incorporated into decision-making. Such retrieval-based architectures underscore the importance of lifelong patient records in developing more robust and context-aware clinical models.

Recognizing that much of EHR data is heterogenous, new models have emerged to tackle the challenge of integrating diverse data types. The Multiple Embedding Model for EHR (MEME) \citep{lee2024multimodal} takes an innovative approach by converting many data sources from the EHR  and turns them into pseudo-notes, effectively mimicking clinical text. This enables pretrained foundation models to process structured EHR data more naturally while preserving categorical relationships. By encoding embeddings separately for each modality, MEME achieves superior feature representation and has demonstrated strong performance in Emergency Department decision-support tasks \citep{chen2023multimodal} across multiple hospital systems. Its success in outperforming traditional machine learning models and even some EHR-specific foundation models suggests that multimodal fusion strategies will play a critical role in the next generation of EHR-based AI.

Meanwhile, generative models are making significant strides in medical research applications. MetaGP \citep{liumetagp}, a 13-billion-parameter generative foundation model, integrates both EHR data and medical literature to provide diagnostic support across a range of clinical scenarios, including rare disease identification and emergency condition management. By training on over ten million EHRs and an extensive corpus of medical literature, MetaGP demonstrates the potential of multimodal generative AI to bridge gaps between clinical practice and biomedical research.

Another innovative approach comes from EHRAgent \citep{shi2024ehragent}, which introduces an LLM-powered agent with a built-in code interface for EHR reasoning. Unlike traditional models that passively predict outcomes, EHRAgent actively generates and executes custom analytical scripts to derive insights from EHR data. This few-shot learning capability enables it to handle complex, multi-tabular reasoning tasks with minimal labeled examples, making it a highly flexible tool for real-world medical problem-solving.

Collectively, these advancements reflect a broader trend in EHR modeling—moving beyond static, one-size-fits-all architectures toward more adaptive, context-aware, and multimodal approaches. Whether through scalable architectures like EHRMamba, temporal models like MOTOR, retrieval-enhanced frameworks, or generative models like MetaGP, 2024 has witnessed a fundamental shift in how AI systems engage with EHR data. These innovations pave the way for more intelligent, real-time, and clinically meaningful AI applications that are better equipped to meet the challenges of modern healthcare.

\section{Data Standards and Evaluations}

In the rapidly evolving field of healthcare AI, the establishment of robust data standards and comprehensive evaluation frameworks is crucial for ensuring the reliability, reproducibility, and applicability of machine learning models in clinical settings.

\subsection{Data Standards}

\paragraph{Medical Event Data Standard (MEDS)}

The question "Do we need data standards in the era of large language models?" is addressed in a study \citep{brat2024we} examining the interplay between LLMs and existing medical data standards. The authors argue that, despite the advanced capabilities of LLMs, standardized data remains crucial for ensuring interoperability, accuracy, and reliability in healthcare applications. They suggest that LLMs should be designed to work within these standards to maintain consistency and trustworthiness in medical data processing. A significant advancement in this area is the introduction of the Medical Event Data Standard (MEDS) \citep{arnrich2024medical}, a lightweight schema designed to facilitate machine learning over electronic health record (EHR) data. Unlike traditional common data models, MEDS offers a minimalistic yet highly interoperable framework that bridges various datasets, tools, and model architectures. By providing a simple standardization layer, MEDS enhances the reproducibility and robustness of machine learning research in healthcare.  

Building upon this foundation, the MEDS Decentralized, Extensible Validation (MEDS-DEV) Benchmark has been developed to establish reproducibility and comparability in machine learning applications for health \citep{kolo2024meds}. MEDS-DEV provides a decentralized framework that allows researchers to validate and compare their models across diverse datasets, promoting transparency and consistency in model evaluation. To streamline the extraction of meaningful cohorts from event-stream datasets, the Automatic Cohort Extraction System (ACES) \citep{xu2024aces} has been introduced. ACES automates the identification of patient cohorts based on specific clinical criteria, thereby accelerating the research process and reducing the potential for human error in cohort selection. Complementing these tools is the Automated Tabularization and Baseline Methods for MEDS (MEDS-Tab) \citep{oufattole2024medstabautomatedtabularizationbaseline}, which focuses on converting complex medical event data into structured tabular formats suitable for analysis. MEDS-Tab provides baseline methodologies for processing MEDS-formatted data, facilitating easier integration with various machine learning pipelines.

\paragraph{Schema-based Standards}

In the realm of schema matching within the EHR space, efforts have been directed towards harmonizing disparate data representations to ensure semantic consistency across systems. By aligning different EHR schemas, researchers can integrate data from multiple sources more effectively, enhancing the comprehensiveness of clinical studies and the generalizability of machine learning models.

EHRmonize \citep{matos2024ehrmonize} is a framework that leverages large language models (LLMs) to extract and abstract medical concepts from electronic health records (EHRs). Using real-world medication data, it evaluates how well LLMs perform on free-text extraction and binary classification tasks across different prompting strategies. The study shows that this approach significantly improves efficiency, cutting annotation time by about 60\%. However, it also emphasizes the need for clinician oversight to ensure accuracy and reliability in real-world clinical settings.

\subsection{Evaluation Frameworks}

\begin{table*}[t!]
    \centering
    \caption{Summary of Benchmarks for Large Language Models in Healthcare}
    \label{tab:evaluation_frameworks}
    \begin{tabular}{|p{3cm}|p{4cm}|p{6cm}|}
        \hline
        \textbf{Benchmark} & \textbf{Key Focus} & \textbf{Evaluation Criteria and Use Cases} \\
        \hline
        \textbf{CliBench} \citep{ma2024clibenchmultifacetedmultigranularevaluation} & Clinical decision-making tasks & Assesses LLMs on diagnoses, procedures, lab test orders, and prescriptions using structured output ontologies \\
        \hline
        \textbf{ClinicalBench} \citep{chen2024clinicalbenchllmsbeattraditional} & Comparison of LLMs with traditional ML models & Evaluates predictive performance of different model architectures across clinical prediction scenarios \\
        \hline
        \textbf{AgentClinic} \citep{schmidgall2024agentclinicmultimodalagentbenchmark} & AI agents in multimodal clinical settings & Evaluates adaptability of AI agents using multimodal data in simulated clinical environments \\
        \hline
        \textbf{EHRNoteQA} \citep{kweon2024ehrnoteqallmbenchmarkrealworld} & Patient-specific question answering & Assesses the ability of LLMs to provide accurate, context-aware answers using patient-specific EHR data \\
        \hline
        \textbf{LongHealth} \citep{adams2024longhealthquestionansweringbenchmark} & Handling long clinical documents & Tests LLMs on extracting and reasoning over detailed medical texts \\
        \hline
    \end{tabular}
\end{table*}

Evaluating the performance of large language models (LLMs) in clinical contexts necessitates multifaceted and granular benchmarks. CliBench \citep{ma2024clibenchmultifacetedmultigranularevaluation} addresses this need by offering a comprehensive evaluation suite that assesses LLMs across various clinical decision-making tasks, including diagnoses, procedures, lab test orders, and prescriptions. By providing structured output ontologies, CliBench enables precise and detailed evaluations, shedding light on the capabilities and limitations of LLMs in healthcare applications. Similarly, ClinicalBench \citep{chen2024clinicalbenchllmsbeattraditional} provides a platform to compare the clinical prediction capacities of LLMs against traditional machine learning models. Encompassing a wide range of models and tasks, ClinicalBench offers insights into the strengths and weaknesses of different modeling approaches in clinical prediction scenarios. 

The Medical Adaptation of Large Language and Vision-Language Models study \citep{jeong2024medicaladaptationlargelanguage} critically examines the progress made in tailoring LLMs and vision-language models for medical applications. By evaluating the adaptations and performance of these models in medical contexts, the study provides valuable insights into their current capabilities and areas needing improvement. Evaluating the performance of predictive models, especially under class imbalance, is critical in healthcare applications. The paper "A Closer Look at AUROC and AUPRC Under Class Imbalance" delved into evaluation metrics commonly used in predictive modeling. The authors analyzed the effectiveness of the Area Under the Receiver Operating Characteristic Curve (AUROC) and the Area Under the Precision-Recall Curve (AUPRC), providing insights into their applicability and limitations. The study emphasized the importance of selecting appropriate metrics to accurately assess model performance in the presence of class imbalance \citep{mcdermott2025closerlookaurocauprc}. In the domain of foundation model representations, the FEET: A Framework for Evaluating Embedding Techniques \citep{lee2024feet} offers a structured approach to assess various embedding methods used in medical machine learning. By providing standardized evaluation metrics, FEET aids researchers in selecting appropriate embedding techniques for their specific applications.

Recognizing the importance of multimodal data in clinical environments, AgentClinic \citep{schmidgall2024agentclinicmultimodalagentbenchmark} introduces a benchmark to evaluate AI agents in simulated clinical settings. By incorporating various data modalities and clinical scenarios, AgentClinic provides a comprehensive platform to assess the performance and adaptability of AI agents in healthcare. To address the need for patient-specific question-answering capabilities, EHRNoteQA \citep{kweon2024ehrnoteqallmbenchmarkrealworld} presents a benchmark designed to evaluate LLMs in clinical settings. By focusing on patient-specific questions, this benchmark assesses the ability of LLMs to provide accurate and relevant information based on individual patient records. Lastly, LongHealth \citep{adams2024longhealthquestionansweringbenchmark} offers a question-answering benchmark that deals with long clinical documents. By challenging models to process and extract relevant information from extensive clinical texts, LongHealth evaluates the proficiency of LLMs in handling complex and detailed medical documents.

\section{Application}

In 2024, the landscape of predictive modeling in healthcare experienced significant advancements \citep{wang2024recentadvancespredictivemodeling}, with researchers leveraging artificial intelligence (AI) to enhance diagnostic accuracy, personalize treatment plans, and improve patient outcomes. A comprehensive survey by Wang et al. systematically reviewed recent developments in deep learning-based predictive models utilizing electronic health records (EHRs). The study categorized various predictive models and highlighted the challenges and future directions in this domain. 

In the realm of chronic disease management, \cite{munirathnam2024artificial} explored the application of AI-powered predictive models for conditions such as diabetes and cardiovascular diseases. Their findings demonstrated that machine learning models, including neural networks and random forests, effectively predict disease progression, enabling earlier interventions and improved risk stratification.  Focusing on chronic kidney disease (CKD), \cite{jawad2025aidrivenpredictiveanalyticsapproach} proposed an AI-driven predictive analytics approach that combines ensemble learning with explainable AI techniques. Their model not only predicts CKD progression but also provides insights into the contributing factors, thereby supporting clinicians in making informed decisions. 

In the context of sepsis, a condition with high mortality rates, \cite{Chang_2024} addressed the need for fairness and transparency in predictive modeling. They introduced a method that enhances model fairness and employs a novel feature importance algorithm to elucidate each feature's contribution to equitable predictions, promoting trust and reliability in clinical applications. In the realm of clinical decision support, research has explored the application of foundation models in prescribing appropriate treatments. The study by \cite{lee2024enhancingantibioticstewardshipusing} examined how these models analyze patient data to recommend antibiotics in line with established clinical guidelines. While the results were promising, the study emphasized the necessity for rigorous validation to ensure safety and efficacy before such models can be reliably implemented in real-world clinical environments. In primary healthcare settings, the implementation of AI-based CDSS has shown promise. A study by \cite{gomez2024artificial} reviewed outcomes of such systems, noting improvements in clinical management, patient satisfaction, and safety, along with reductions in physician workload. However, the study also highlighted challenges related to physician perceptions and cultural settings, suggesting that further research is needed to optimize AI-CDSS applications in diverse clinical environments.

The integration of AI into clinical trials has also been a focal point, with  recent work \citep{Sydney_Anuyah_2024, wornow2025zero, jin2024matching}. exploring how deep learning and predictive modeling can optimize trial design, patient recruitment, and real-time monitoring. Their study highlighted the potential of AI to stratify patients, forecast adverse events, and personalize treatment plans, thereby bridging precision medicine and patient-centered care. \cite{yu2025healthllmpersonalizedretrievalaugmenteddisease} introduced Health-LLM, a personalized retrieval-augmented disease prediction system that combines large-scale feature extraction with medical knowledge trade-off scoring. This framework integrates health reports and medical knowledge into a large language model, enhancing disease prediction accuracy and supporting personalized health management. 

The role of LLMs in medical coding has also been critically evaluated. The paper Large Language Models are Poor Medical Coders—Benchmarking of Medical Code Querying by \cite{soroush2024large} assessed the performance of LLMs in generating correct medical codes from clinical descriptions. Despite their advanced language capabilities, the models demonstrated limited accuracy in this task. This finding points to inherent challenges LLMs face in understanding and applying the structured language of medical coding, suggesting that reliance on these models for automated coding may be premature without further refinement. One area of investigation centers on the proficiency of LLMs in interpreting medical codes. A study by \cite{lee2024largelanguagemodelsabstract} delved into this by assessing the ability of LLMs to accurately map alphanumeric codes, such as ICD-10, to their corresponding medical terminologies. The findings revealed that current LLMs face challenges in this domain, often struggling to establish precise mappings. This underscores a significant gap in their comprehension of structured medical coding systems, highlighting the need for enhanced representations within LLMs to improve their utility in clinical settings.

Beyond coding, LLMs have been applied to specific areas of healthcare, such as maternal health. The paper "NLP for Maternal Healthcare: Perspectives and Guiding Principles in the Age of LLMs"  by \cite{antoniak2024nlp} discussed the application of Natural Language Processing techniques, powered by LLMs, in analyzing and interpreting data related to maternal health. The authors outlined potential applications and ethical considerations, advocating for the development of guidelines to ensure responsible and effective use of LLMs in this sensitive area of healthcare. Patient behavior analysis has also benefited from LLM applications. In the study \cite{miao2024identifying} researchers employed LLMs to analyze patient narratives, aiming to identify factors contributing to changes in contraceptive methods. The findings demonstrated the potential of LLMs to extract meaningful insights from unstructured data, thereby aiding in understanding patient behaviors and informing healthcare strategies.

\section{Conclusion}

The rapid advancements in Artificial Intelligence, particularly in machine learning and large language models, are profoundly reshaping the landscape of healthcare. This paper has highlighted the transformative potential of these technologies in enhancing clinical decision-making, diagnostics, and patient care. However, the pace of AI innovation poses significant challenges to the traditional academic publishing model, which often lags behind, rendering scholarly contributions quickly obsolete.

To address this gap, we advocate for the adoption of an annualized citation framework that emphasizes the most recent AI-driven healthcare innovations. By systematically referencing the latest breakthroughs, this approach ensures that research remains current and relevant, fostering a more dynamic and informed academic discourse. Such a framework not only enhances the immediacy and applicability of AI research in healthcare but also aligns scholarly communication with the fast-evolving nature of the field.

Our exploration of recent developments in language models and Electronic Health Record (EHR) foundation models underscores the necessity for adaptable and scalable AI systems tailored to the complexities of medical data. Innovations like MediSwift, BMRetriever, and EHRMamba exemplify the strides being made towards more efficient and accurate healthcare AI applications. Additionally, the establishment of robust data standards and comprehensive evaluation frameworks, as discussed, is crucial for ensuring the reliability and reproducibility of AI models in clinical settings.

The applications of AI in predictive modeling further demonstrate its potential to revolutionize patient care through improved diagnostic accuracy, personalized treatment plans, and enhanced patient outcomes. Nonetheless, challenges such as model fairness, transparency, and the integration of AI systems into existing clinical workflows must be meticulously addressed to realize the full benefits of these technologies.

Looking forward, the proposed annualized citation framework represents a pivotal step towards synchronizing academic research with the swift advancements in AI. By embracing this innovative publication model, the healthcare AI community can ensure that research remains at the forefront of technological progress, ultimately driving more effective and timely solutions in patient care. Continued collaboration between researchers, clinicians, and policymakers will be essential in navigating the complexities of AI integration, fostering an environment where cutting-edge innovations can thrive and translate into meaningful clinical impact.

\bibliography{iclr2025_conference}
\bibliographystyle{iclr2025_conference}

\end{document}